\documentclass[%
 aip,%
 jcp,%
 amsmath,%
 amssymb,%
 reprint]{revtex4-1}
\usepackage[cm]{fullpage}
\usepackage{float}
\usepackage{graphicx}
\usepackage{bm}

\begin{document}
\newcommand{\sub}[1]{_{\mbox{\scriptsize {#1}}}}

\preprint{AIP/123-QED}

\title[]{Properties of Liquid Clusters in Large-scale Molecular Dynamics Nucleation Simulations}
\author{Raymond Ang\'elil}
  \affiliation{Institute for Theoretical Physics, University of Zurich, 8057 Zurich, Switzerland}

\author{J\"urg Diemand}
  \affiliation{Institute for Theoretical Physics, University of Zurich, 8057 Zurich, Switzerland}

\author{Kyoko K. Tanaka}
  \affiliation{Institute of Low Temperature Science, Hokkaido University, Sapporo 060-0819, Japan}

\author{Hidekazu Tanaka}
  \affiliation{Institute of Low Temperature Science, Hokkaido University, Sapporo 060-0819, Japan}

\date{\today}
\begin{abstract}
We have performed large-scale Lennard-Jones molecular dynamics simulations of homogeneous vapor-to-liquid nucleation, with $10^9$ atoms. This large number allows us to resolve extremely low nucleation rates, and also provides excellent statistics for cluster properties over a wide range of cluster sizes. The nucleation rates, cluster growth rates, and size distributions are presented in Diemand et al. [J. Chem. Phys. {\bf 139}, 74309 (2013)], while this paper analyses the properties of the clusters. We explore the cluster temperatures, density profiles, potential energies and shapes. A thorough understanding of the properties of the clusters is crucial to the formulation of nucleation models. Significant latent heat is retained by stable clusters, by as much as $\Delta kT = 0.1 \epsilon$ for clusters with size $i = 100$. We find that the clusters deviate remarkably from spherical - with ellipsoidal axis ratios for critical cluster sizes typically within $b/c = 0.7\pm 0.05$ and $a/c = 0.5 \pm 0.05$. We examine cluster spin angular momentum, and find that it plays a negligible role in the cluster dynamics. The interfaces of large, stable clusters are thiner than planar equilibrium interfaces by $10-30\%$. At the critical cluster size, the cluster central densities are between $5-30\%$ lower than the bulk liquid expectations. These lower densities imply larger-than-expected surface areas, which increase the energy cost to form a surface, which lowers nucleation rates. 
\end{abstract}

\maketitle

\section{\label{sec:level1}Introduction\protect\\}

A homogeneous vapor, when supersaturated, changes phase to liquid through the process of nucleation. This transformation is stochastically driven through the erratic formation of clusters, made up of atoms clinging together in droplets large enough that the free energy barrier is surpassed. Despite the ubiquity of this process in nature, attempts to model this process have met with difficulty, because the properties of the nanoscale liquid-like clusters are not well known \cite{KalikmanovBook}. Unlike real laboratory experiments (Sinha, S. et al. (2010)\cite{sinha} for example), computer simulations offer detailed information on the properties and evolution of clusters. However, direct simulations of nucleation are typically performed using a few thousand atoms only, and are therefore limited to extremely high nucleation rates and to forming a small number stable clusters (e.g. Wedekind et al. (2007)\cite{Wedekind2007} and Napari et al. (2009)\citep{Napari2009}). A large number of small simulations allows us to constrain the critical cluster properties in the high nucleation rate regime, but there seems to be limited information in the literature on critical cluster properties.
An alternative approach is to simulate clusters in equilibrium with a surrounding vapor, however the resulting cluster properties seem to differ significantly from those seen in nucleation simulations \cite{Napari2009}.

We report here on the properties of the clusters which form in large-scale molecular dynamics Lennard-Jones simulations. These direct simulations of homogeneous nucleation are much larger and probe much lower nucleation rates than any previous direct nucleation simulation. Some of the simulations even cover the same temperatures, pressures, supersaturations and nucleation rates as the recent Argon supersonic nozzle (SSN) experiment\cite{sinha} and allow, for the first time, direct comparisons to be made. The nine simulations we analyse here are part of a larger suite of runs: results of these runs pertaining to nucleation rates and comparisons to nucleation models and the SSN experiment\cite{sinha}, are presented in Diemand et al. (2013)\cite{paper1}. 

The large size of these simulations is primarily necessitated by the rarity of nucleation events at these low supersaturations. However, a further benefit gained from large simulations is the substantial number of nucleated droplets which are able to continue growing without significant decreases in the vapor density. This allows us to study, with good statistics, the properties of clusters as they grow, embedded within an realistic unchanging environment. This is particularly important in understanding the role that the droplet's surface plays in the development of the droplet - as a bustling interface between the denser (and ever-growing) core, and the vapor outside. The nucleation properties of the simulations, the cluster growth rates and size distributions, and comparisons to nucleation models are presented in Diemand et al. (2013)\cite{paper1}, while here we explore the properties of the clusters themselves. Studying the properties of the nano-sized liquid clusters which form, both stable and unstable, is of service to understanding the details of the nucleation process, the reasons behind the shortcomings of the available nucleation models, and aids in the blueprinting and selection of ingredients for future ones.

Section II provides details on the numerics of the simulations. In Section \ref{sec:temperatures} we present the temperatures of the clusters and in
Section \ref{sec:potential} we show the clusters' potential energies. Section \ref{sec:rotation} addresses cluster rotation and angular momentum. 
The shapes the clusters take on is detailed in section \ref{sec:shapes}.
The cluster density profiles are explored in section \ref{sec:densities}.
Section \ref{sec:sizes} will address cluster sizes and we use this information to revisit nucleation theory in section \ref{sec:revisiting}.

\section{Numerical Simulations}
The simulations were performed with the Large-scale Atomic/Molecular Massively Parallel Simulator \cite{lammps}, an open-source code developed at Sandia National Laboratories. The interaction potential is Lennard-Jones,

\begin{equation}\label{eqn:pot}
\frac{U\left(r\right)}{4\epsilon} = \left(\frac{\sigma}{r}\right)^{12} - \left(\frac{\sigma}{r} \right)^6,
\end{equation}
though truncated at $5\sigma$, as well as shifted to zero. The properties of the Lennard Jones fluid depend on the cutoff scale, and our chosen cutoff is widely used in simulations, resulting in properties similar to Argon at reasonable computational cost. The integration routine is the well-known Verlet integrator (often referred to as leap-frog), with a time-step of $\Delta t = 0.01\tau = 0.01\sigma\sqrt{m/\epsilon}$, regardless of the simulation temperature. In the Argon system, the units become $\sigma = 3.405\,\text{\AA}$, $\epsilon/k = 119.8\,\text{K}$, and $\tau = 2.16\,\text{ps}.$

The initial conditions correspond to random\cite{random_generator} positions, and random velocities corresponding to a chosen temperature, in a cube with periodic boundary conditions. The simulations are carried out at constant average temperature, but not constant energy. The velocities of all the atoms in the simulation are rescaled, so that the average temperature of the simulation box remains constant\cite{nonisothermal}. The condensation process converts potential energy into kinetic energy as the clusters form, adding heat to the clusters. However because the amount of atoms in gas relative to the number in the condense phase is so large, the amount of thermostatting required is negligible.  Because the nucleation rates explored in these simulations are relatively low, relative to the number of monomers in the simulation, very few clusters form. This means that the average heat produced through the transformation is extremely low, and so the amount of velocity rescaling necessary almost negligible. 

\begin{table*}[]
\caption{Simulation properties: temperature $T$, atom initial number density $n_{t=0}$, total run time $t_{\rm{end}}$, supersaturation $S$, nucleation rate $J_{\textrm{MD}}$, and estimates for the critical cluster size $i^{*}$. }
\begin{ruledtabular}
\begin{tabular}{ l c c c c c c}
Run ID&T& $n_{t=0}$& $t_{\rm{end}}$  &  $S$&$J_{\textrm{MD}}$\cite{paper1} & $i^*$\cite{paper1} \\
& [$\epsilon/k$] &   $\left[\sigma^{-3}\right]$ & $\left[\tau \right]$ & &$\left[ \sigma^{-3}\tau^{-1}\right]$
\\
\hline
  T10n60 & 1.0 &   $6.00\times10^{-2}$ & $2.55 \times 10^{3}$ & $1.63$&$7.93\pm \times 10^{-13}$ & $126$ \\
   
  T10n55 & 1.0 &  $5.50\times10^{-2}$ & $2.37 \times 10^{4}$ & $1.55$&$<1.1\times 10^{-14}$ & $>126$ \\ \hline
  
  T8n30 & 0.8 &   $3.00\times10^{-2}$ & $3.98 \times 10^{3}$ & $4.02$&$2.57\pm 0.02 \times 10^{-10}$ & $48$\\ \hline
  
  T6n80 & 0.6 & $8.00\times10^{-3}$ & $5.00 \times 10^{3}$ & $16.9$&$1.09\pm0.01\times 10^{-12} $ & $24$ \\

  T6n65 & 0.6 &   $6.50\times10^{-3}$ & $3.00 \times 10^{4}$ & $14.0$&$2.58\pm0.19\times 10^{-15}$ &$38$ \\
  
  T6n55 & 0.6 & $5.00\times10^{-3}$ & $1.81 \times 10^{5}$ & $11.95$&$0.49-7.21\times 10^{-17}$ & $21-40 $ \\ \hline

  T5n40 & 0.5 &  $4.00\times10^{-3}$ & $4.20 \times 10^{3}$ & $72.8$&$2.74\pm 0.14\times10^{-12}$ &$18$ \\ \hline

  T4n10 & 0.4 & $1.00\times10^{-3}$ & $3.95 \times 10^{4}$ & $484$&$1.49\pm0.01\times10^{-14}$ & $12$\\
 
  T4n7 & 0.4 & $0.70\times10^{-3}$ & $2.85 \times 10^{5}$ & $342$&$8.99\pm0.3\times10^{-17}$ & $14$\\
 
\end{tabular}
\end{ruledtabular}\label{tab:t1}
\end{table*}

The initial density of the simulations are indicated in table \ref{tab:t1}. The monomer density drops rapidly as the gas subcritical cluster distribution forms. The clusters are identified by the Stillinger \cite{Stillinger} algorithm, whereby neighbours with small enough separations are joined into a common group. The linking length is temperature dependent, and given in table \ref{tab:t2}. It is set by the distance below which two monomers would be bound in the potential (\ref{eqn:pot}), were their velocities to correspond to the thermodynamic average, $v^2 = 3kT$\citep{linking, tanaka2011}.
The type of initial conditions, the simulation time-step,  thermostatting consequences, and the Lennard-Jones cutoff, have been investigated through convergence tests in Diemand et al. (2013)\cite{paper1}. 

\begin{figure}[]
\includegraphics[scale = 0.175]{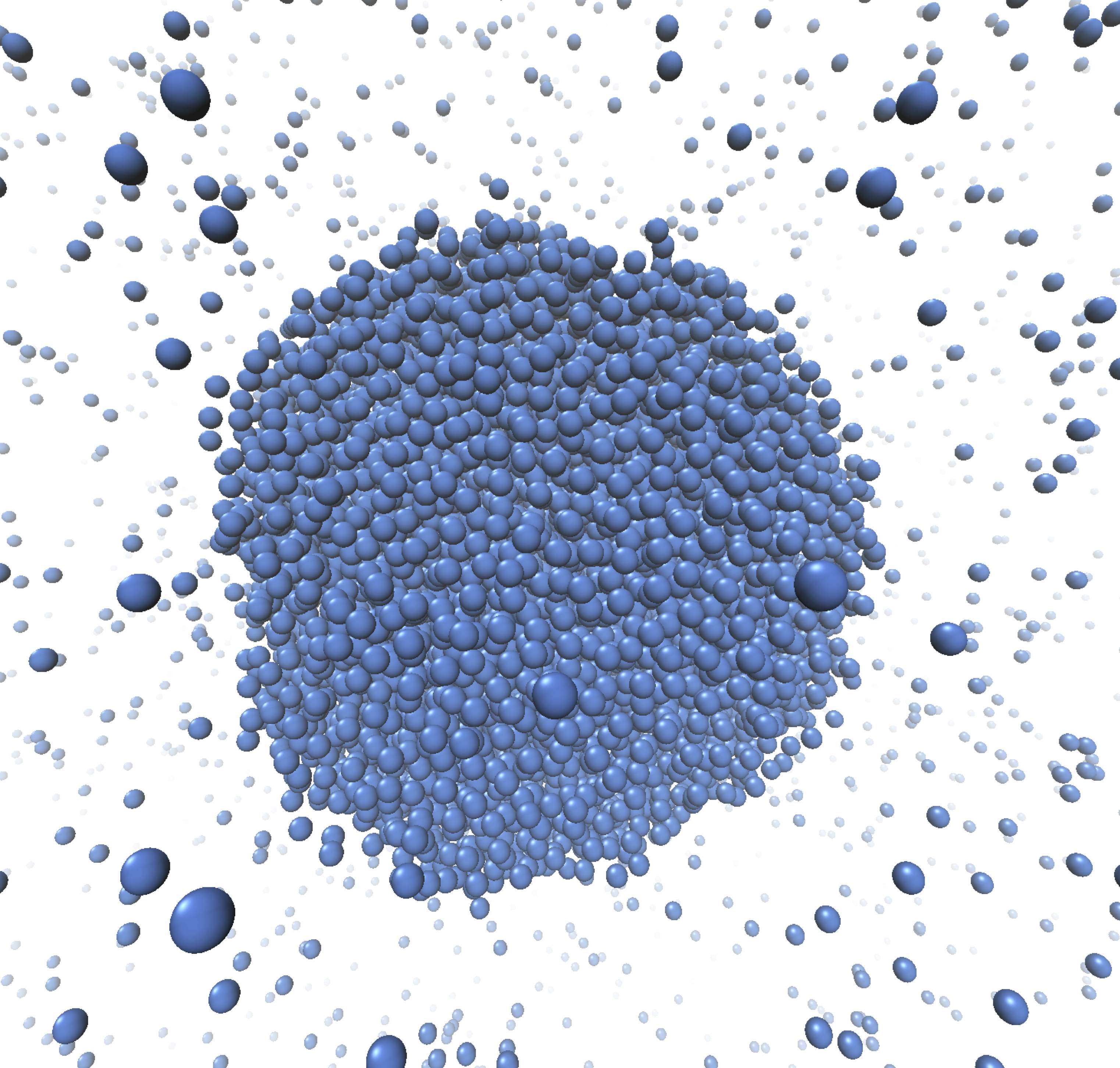}
\caption{The only stable cluster which formed in run T6n55. It is made up of 7373 members, as identified by the Stillinger search criterion. Its density and potential energy profiles are shown in figure \ref{fig:big_cluster_density}. Nucleation theory reasons that clusters are spherical, yet we often observe deformation, especially for small clusters, as well as at high temperatures.}\label{fig:w00t}
\end{figure}

The runs we have chosen to analyse here cover a broad temperature range $0.4$-$1.0 \epsilon/k$, and a nucleation rate range $\sim 10^{-17}$ - $ 10^{-10} \sigma^{-3}\tau^{-1}$. All the results presented in this paper come from single snapshots taken at the end of each run. Because of the different nucleation rates, cluster growth rates, critical cluster sizes, and because all runs were terminated at a different time, different runs have markedly different size distributions at the time of termination. For example, while run T8n3 formed $>10^5$ stable clusters, the largest of which has $>2500$ members, run T6n55 formed only a single stable cluster, yet made up of $\sim 10^4$ atoms (pictured in figure \ref{fig:w00t}), while T10n55 formed clusters up to $\sim 100$, although still below the critical size for this run. However, even a run which forms no stable clusters is still of value, as the subcritical droplets, whose properties we are interested in, are always present in the gas. 
Sometimes it is convenient to group the atoms of each cluster into one of two categories: according to whether they belong to the core, or to the surface. Atoms are considered core atoms if they have at least $n_{core}$ neighbours within the search radius. $n_{core}$ is temperature dependent, and is chosen such that the number of neighbours at the bulk density is approximately constant. The cutoffs are given in table \ref{tab:t2}. Unless otherwise indicated, error bars indicate the root mean square scatter in the measured quantity. In many cases, it is instructive to make comparisons between the liquid in the clusters of various sizes (and therefore curvatures), and the bulk liquid. To facilitate this comparison, we perform supplementary MD saturation simulations of the vapor-liquid phase equilibrium\citep{slabz, baidakov, Trokhymchuk, Mecke1997}, in order to calculate the relevant quantities directly. Appendix \ref{bulk} details these simulations, whose parameter-value results are given in table \ref{tab:bulksims}.

\begin{table}[]
\caption{Thermodynamic quantities and simulation parameters at each temperature. Pressures at saturation $P_{\rm sat}$ are taken from Trokhymchuk et al.\cite{Trokhymchuk}, which agrees well with our equilibrium simulations, see Appendix \ref{bulk}.
}
\begin{ruledtabular}
\begin{tabular}{ l c c c} $T$ &    
 $P_{\rm sat}$ & 

 $r\sub{c}$ &
 $n_{\textrm{core}}$
\\ $[\epsilon/k]$& $ [\epsilon/\sigma^3]$&  $[\sigma]$ 
\\  \hline     
 1.0  &$2.55 \times 10^{-2}$  
     &  1.26 & 4 \\ \hline
 0.8  &$4.53 \times 10^{-3}$   
    &  1.33 & 5 \\\hline
 0.6    &$2.54 \times 10^{-4}$   
 & 1.41 & 7 \\\hline
 0.5    &$2.54 \times 10^{-5}$   
 & 1.46 &8 \\\hline
 0.4  & $8.02\times 10^{-7}$    
 &  1.52 &8 \\
\end{tabular}
\end{ruledtabular}\label{tab:t2}
\end{table}

\section{Temperatures}\label{sec:temperatures}

\begin{figure}[]
\includegraphics[scale = 0.6]{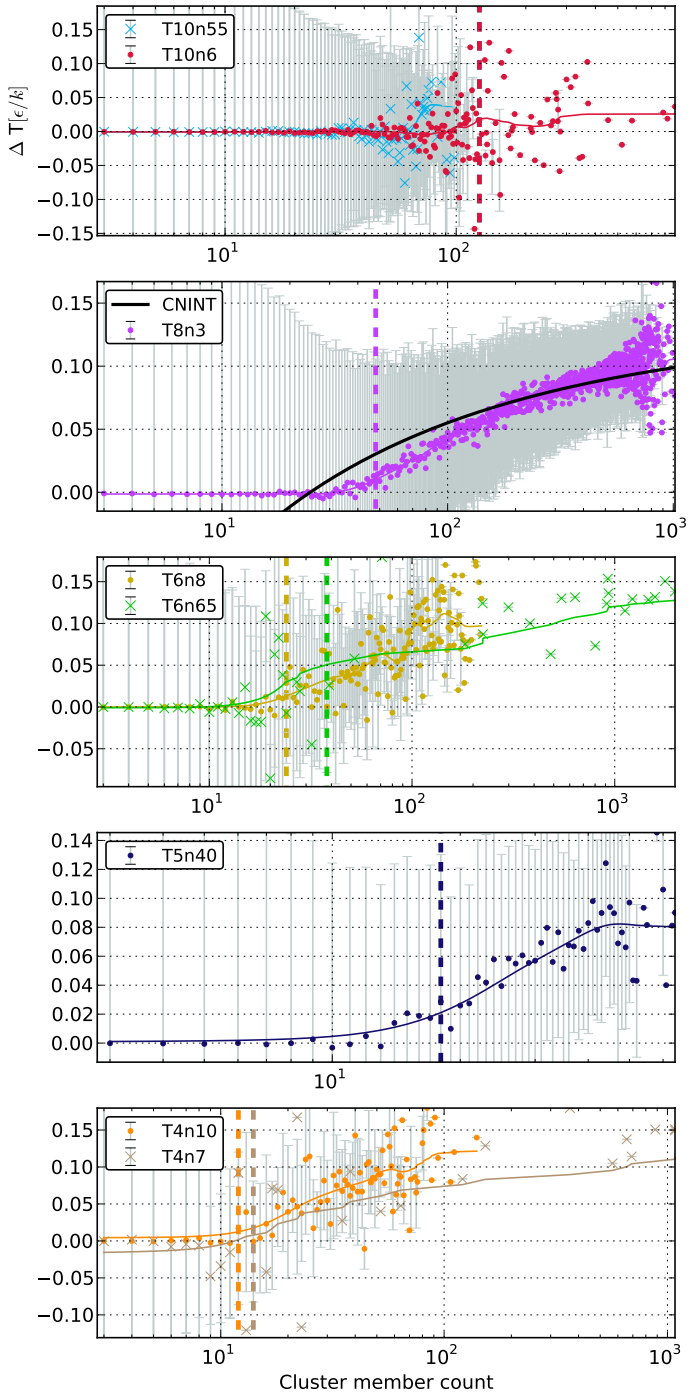}
\caption{The difference between the cluster temperature and the simulation average. The grey bars indicate the r.m.s. scatter. The solid lines indicate a sliding average over a window of size 12 in the cluster member count. The vertical dashed lines indicate the estimated critical cluster sizes, using the first nucleation theorem. Only stable clusters retain latent heat. At all temperatures except the highest ($kT=1.0\epsilon$), the clusters are hotter than the gas due to the latent heat of formation, which has not yet had enough time to dissipate back into the gas. 
The thick solid line in the second panel shows the predicted $\Delta T_C = T_C - T_0$ from CNINT\citep{feder}, note that $\Delta T_C < \Delta T$ for small cluster and $\Delta T_C \simeq \Delta T$ for cluster larger than about 50 atoms, see for example Wedekind et al. (2007)\cite{nonisothermal}.
}\label{fig:cluster_temperatures}
\end{figure}

We define the temperature of an ensemble of atoms from their mean kinetic energy,  
\begin{equation}
kT \equiv  \frac{2}{3}\left< E_{\textrm{kinetic}}\right> = \frac{1}{3N} \sum_{i=1}^N m v_i^2,
\end{equation}
where $N$ is the total number of atoms in the ensemble, $m_i$ are the atom masses, and the velocities $v_i$ are those relative to the simulation box. For small, out-of-equilibrium clusters it can be troublesome to define temperature as an average over kinetic energy. However, by taking ensemble averages, the large number of small, subcritical clusters in the simulations mitigate this complication. 
The cluster temperature differences with the average bath temperature, $\Delta T$ is plotted for a few runs in figure \ref{fig:cluster_temperatures}, and shows that clusters smaller and at the critical cluster size are in thermodynamic equilibrium with the gas. This can be expected as sub-critical clusters are as likely to accrete a monomer as they are to evaporate one: Because the growth rate is equal to the loss rate, subcritical clusters quickly lose latent heat due to the many interactions that they undergo as they random-walk the size ladder. Wedekind et al. 2007 \cite{nonisothermal} find a similar behaviour
in $T=0.4 \epsilon / k$ simulations. Their temperature differences are larger and set in at smaller cluster sizes than those in our $T=0.4 \epsilon / k$ simulations, as expected at their much higher nucleation rates.

Larger than the critical size, the cluster temperatures relative to the gas temperature increase with the cluster size. For simulations at the same temperature, the higher supersaturation case has a higher latent heat retention. This is likely due to the higher growth rates, caused by the higher collision rate and also by
the higher probability that a monomer sticks to a cluster\cite{tanaka2011,paper1}.

The only set of runs without a significant post-critical $\Delta T$ signal are the high temperature $kT = 1.0\epsilon$ runs. It is possible that evaporation, which is proportional to $1/S\;$ \cite{paper1}, is efficient enough at the low supersaturations of the $kT = 1.0\epsilon$ runs to keep the clusters closer to thermal equilibrium with the gas.

If we divide the atoms of every cluster into two population types: core atoms and surface atoms, based on the number of neighbours that each atom has, we can investigate their temperature differences. Across all of our simulations, we find no significant difference in the core temperatures vs. the surface temperatures. The clusters are conductive enough for the surfaces to maintain thermodynamic equilibrium with their cores. 

\begin{figure}[]
\includegraphics[scale = 0.6]{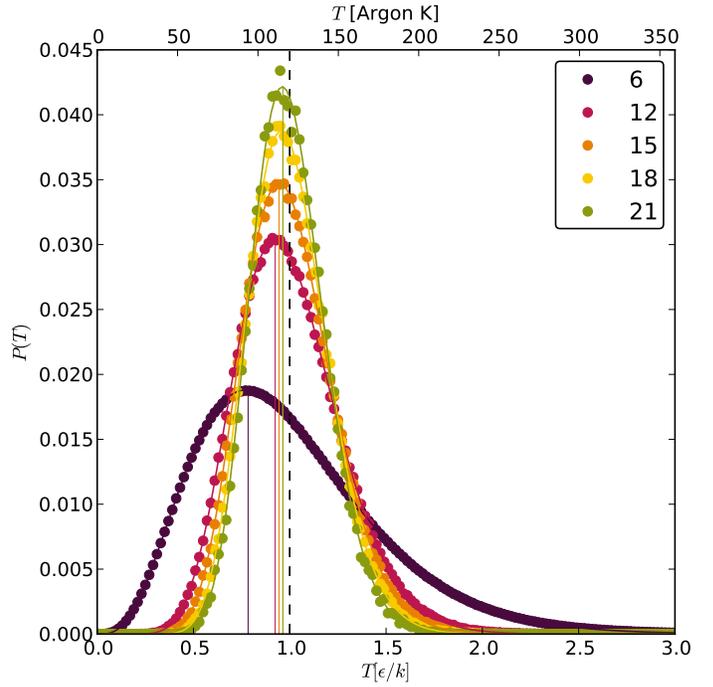}
\caption{Directly measured temperature probability distributions (circles) for a few cluster sizes for the run T10n6. All clusters below the critical size have average temperature (the dotted vertical line) equal to the gas temperature (see figure \ref{fig:cluster_temperatures}), however the most probable temperature $T_C$ (solid vertical lines), is lower for small clusters. The theoretical prediction from equation (\ref{david}) from McGraw and Laviolette\cite{mcgraw} fits (solid, curved lines) these measured distributions extremely well.
}\label{fig:temp_dist_one_run}
\end{figure}

\begin{figure}[]
\includegraphics[scale = 0.6]{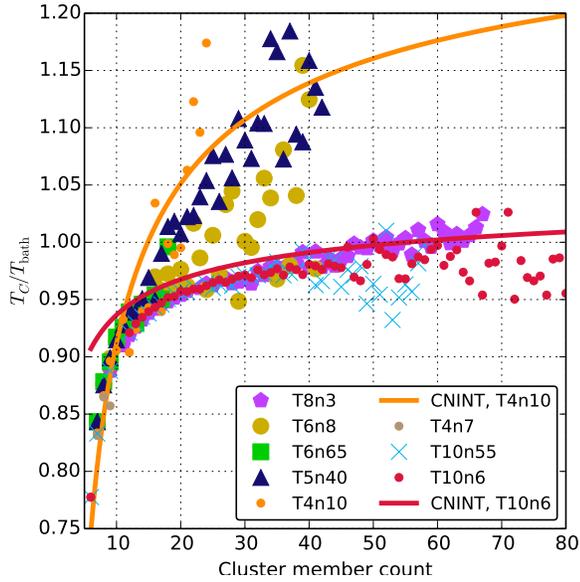}
\caption{The ratio of the most probable cluster temperatures, $T_C$,  from fits to equation (\ref{david}), to the gas temperature. The solid lines indicate the predictions from CNINT from Feder et al. 1966\citep{feder} for runs T4n10 and T10n6.}\label{fig:TCs}
\end{figure}

Figure \ref{fig:temp_dist_one_run} shows the temperature probability distribution for clusters of various sizes, for the high-temperature run T10n6. Due to the asymmetry of the distribution for smaller clusters, the average temperature $T_0$ is not equal to the most probable one, $T_C$. The distributions as a function of cluster size were derived by McGraw and Laviolette \cite{mcgraw}:
\begin{equation}\label{david}
P\left( T\right) = K \exp\left[C_v \left(T_C - T \right) + C_v T_C\ln\left(\frac{T_C}{T} \right) \right],
\end{equation}
where $K$ provides normalisation,  and $C_v$ is the cluster's heat capacity. This predicted form fits the distributions shown in Figure \ref{fig:temp_dist_one_run} extremely well and allows us to derive the most probable temperature $T_C$ very accurately. 
We fit this curve to the temperature probability distributions to all runs and cluster sizes where we have sufficient statistics and plot the resulting $T_C$ values in figure \ref{fig:TCs}. Figures \ref{fig:cluster_temperatures} and \ref{fig:TCs} also show comparisons with Feder et al (1966)'s\citep{feder} classical non-isothermal nucleation theory (CNINT), assuming a sticking probability of one, no carrier gas and no evaporation. According to this theory $\Delta T_C = T_C - T_0$ is negative below the critical cluster size, zero for critical clusters and positive above the critical size. The CNINT agrees only qualitatively with the MD results. Discrepencies are expected because the classical nucleation theory does not match the critical cluster sizes, size distributions and nucleation rates found in our MD simulations \cite{paper1}. Similar qualitative agreement was found in $kT = 0.4\epsilon$ simulations with much higher nucleation rates in\cite{nonisothermal}.

For the heat capacity $C_v/k$, all simulations are consistent with a simple linear fit with slope of $3/2$ against $kT$, i.e. the heat capacity per molecule is almost equal to the ideal value for a monoatomic gas.

We remark that: 
\begin{itemize}
\item For clusters larger than the critical cluster size, significant latent heat is retained by the clusters. An exception to this are the high temperatures runs, for which temperatures of clusters greater than the critical size are consistent with zero (within the scatter) - no clear latent heat retention signal is observed. The excess temperature $\Delta T$ is marginally larger for the lower gas temperature ($kT_{\textrm{gas}} = 0.4\epsilon$) runs than the higher ($kT_{\textrm{gas}} = 0.8\epsilon$). 
\item Within the scatter, the cluster core temperatures are the same as the surface temperatures. Heat is conducted efficiently enough that latent heat, which is added to the surface is transferred to the core rapidly enough to maintain equilibrium - within the bounds of statistical veracity.
\item The temperature probability distributions as a function of cluster size have the expected form - reshaping from the Maxwell distribution towards the normal distribution as cluster size grows. 
\item For clusters smaller than the critical size, the most probable temperature $T_C$ relative to the gas temperature $T_\textrm{bath}$ is universal across all runs.  
\end{itemize}

\section{Potential Energies}\label{sec:potential}

\begin{figure}[]
\includegraphics[scale = 0.6]{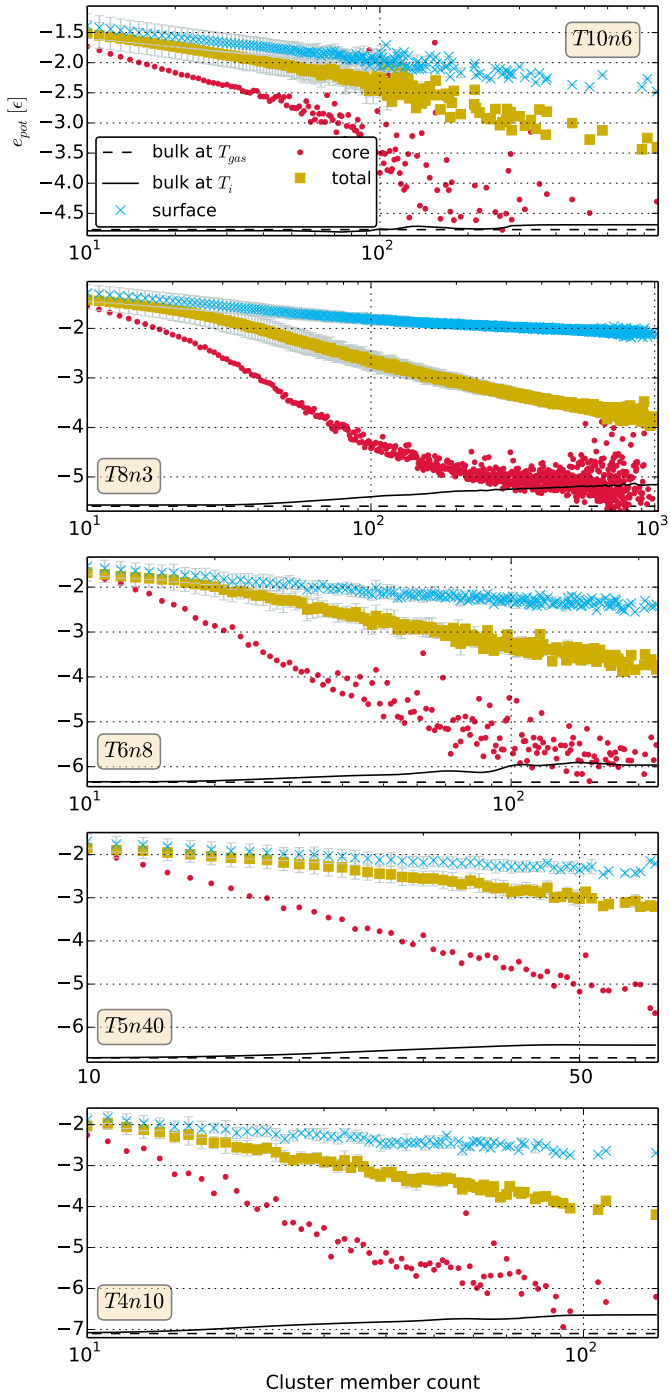}
\caption{Potential energies per particle split into the two population types: core and surface. The round red markers indicate the potential energies of the core particles, and the crosses indicate those of the surface particles. Squares correspond to the total per-particle potential. The horizontal black lines indicate the bulk potential energies from supplementary simulations at the gas target temperature. See appendix \ref{bulk}.}\label{fig:core_vs_surface_PE}
\end{figure}

In atomistic theories of nucleation, e.g. the Fisher droplet model\cite{Fisher1967} and other atomistic models (see e.g. Kashchiev (2000)\citep{KashchievBook} and Kalikmanov (2013)\citep{KalikmanovBook} for details), one relates the surface energy of a cluster $W(i)$ to its total potential energy $E_{\rm pot}(i)$:
\begin{equation}\label{Wi}
E_{\rm pot}(i) = i e_{\rm pot,l}  +  W(i) \; ,
\end{equation}
where $e_{\rm pot,l}$ is the potential energy per particle in the bulk liquid phase. In this approach the surface energy of a cluster is simply the difference between its actual potential energy and the one it would have if all its members were embedded in bulk liquid. We have tried this approach using the bulk liquid potential energies measured in the equilibrium simulations described in the Appendix and find that the resulting surface terms are too large: In the atomistic model the free energy difference at saturation (i.e. where the volume term vanishes) are equal to the surface term, with a constant shift to have zero for monomers:
\begin{equation}
\Delta G_{\rm atomistic}(i,S=1) = W(i) - W(1) \; .
\end{equation}
For $S \neq 1$, the classical volume term is no longer null: 
\begin{equation}
\Delta G_{\rm atomistic}(i) = - (i-1) \ln(S) + W(i) - W(1).
\end{equation}
 This estimate for the free energy lies far above the true free energy landscape, which we reconstructed from the size distribution in the simulation using a new analysis method (Tanaka, K. et al (2014), in preparation). Figure \ref{fig:delG_it8n3} plots this reconstruction	.   In comparison, the atomistic theory free energy curve lies far above these estimates, reaching a critical size $i^* = 130$, at which point the atomistic free energy is $\sim 80 \epsilon$.  The same was found in other simulations, i.e. this simple implementation of an atomistic model seems to overestimate the surface energy and therefore underestimates the nucleation rates by large factors. On the other hand, if not the bulk liquid phase value is chosen for $e_{\rm pot, i}$, but the core potential energy at size $i$ (red dots on figure \ref{fig:core_vs_surface_PE}), then the corresponding surface energy is too low.

\begin{figure}[]
\includegraphics[scale = 0.6]{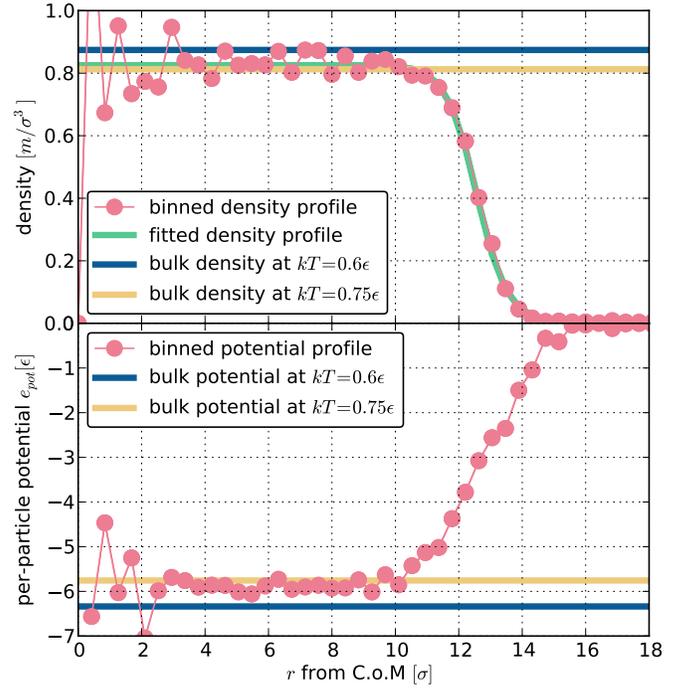}
\caption{The upper panel shows the binned center-of-mass density profile of the large cluster pictured in figure \ref{fig:w00t}. Compared to the gas, this cluster has a temperature excess of $\Delta kT = 0.15\epsilon$. The fit (in green) to equation (\ref{chapela}) puts the central density at $\rho_l = 0.825 m/\sigma^3 $, the midpoint of transition region at $R = 12.56\sigma,$ and the length of the transition region $d = 1.90 \sigma.$ Bulk simulations at $kT = 0.75\epsilon$ have $d = 2.15\sigma.$ Its inner density agrees (within the scatter), with the bulk case at this higher temperature. The lower panel shows its per-particle potential energy, consistent with what the bulk at the raised temperature expects.  }\label{fig:big_cluster_density}
\end{figure}

The lower panel of figure \ref{fig:big_cluster_density} shows the potential energy per particle as a function of distance from the center-of-mass for the large cluster pictured in figure \ref{fig:w00t}. It reaches the same values as in the bulk-liquid at the same temperature. While we observe the kinetic energy differences (i.e. temperature differences, see Section \ref{sec:temperatures}) between core and surface atoms to be consistent with zero, the potential energies of the core are expected to be considerably lower than those of the surface particles, as they have more neighbours. Figure \ref{fig:core_vs_surface_PE} plots the potential energies of core and surface atoms for a few runs. As the clusters grow, they become more tightly packed, and so both the surface and core potential energies drop. The potential energies per core particle are expected to reach a minimum in the limit that the clusters grow large enough to have core potential energies equal to the bulk liquid. In our low-temperature simulations, this occurs at $i\sim 100$, and for our high temperatures simulations at $i\sim 500$. 

To predict nucleation rates accurately one needs an good estimate of the surface term for clusters near the critical size. Figure \ref{fig:core_vs_surface_PE} shows that the core particles in these relatively small clusters are still strongly affected by the surface and the transition region, i.e. their potential energies per particle are far less negative than those found for true bulk liquid particles. This discrepancy might be related to the failure of the simple atomistic model described above. Replacing $e_{\rm pot,l}$ in Eqn. \ref{Wi} with the less negative values actually found in the cores of critical clusters can significantly improve the $\Delta G_{\rm atomistic}$ estimates, at least for $i \simeq i_{\rm crit}$, and lead to better nucleation rate estimates. 

\pagebreak
\newpage

\section{Rotation}\label{sec:rotation}

\begin{figure}
\includegraphics[scale=0.6]{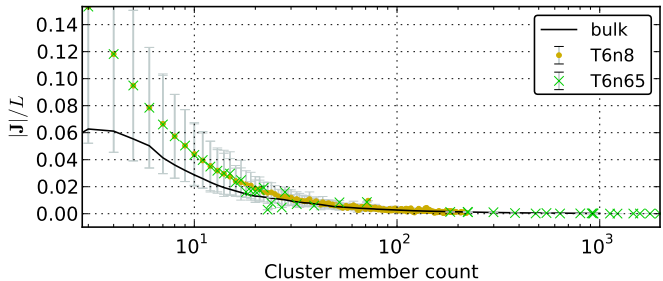}
\caption{This ratio (see definitions (\ref{ang1}) and (\ref{ang2})) is an indicator of the extent to which the members of a cluster rotate in unison. Unity corresponds to harmonious rotation, while zero to dissonance. The cluster spin damps rapidly as they grow. While we plot this quantity for only two runs here, it exhibits similar behavior for all runs. For comparison, this ratio was calculated in the bulk to determine the noise level, as this signal is expected to average to zero for the bulk. Finite-size effects however contribute an intrinsic noise to this quantity. We estimate this by randomly centering spheres in the bulk, and calculating the ratio for the enclosed atoms.}\label{fig:L_over_L_tilde}
\end{figure}

Because clusters grow through isotropic interactions with vapor atoms, it can be expected that the spin of the clusters decreases for larger clusters. 
The angular momentum of a single particle $i$ in a cluster is 
\begin{equation}
\mathbf{j}_i \equiv  \mathbf{r}_{i,{\textrm{C.o.M}}} \times \mathbf{v}_{i,\textrm{C.o.M}},
\end{equation}
where $\textrm{C.o.M}$ denotes that the quantity is taken relative to the cluster center-of-mass, and we have set the mass to unity. The magnitude of the total angular momentum of the cluster is then
\begin{equation}\label{ang1}
\left|\mathbf{J}\right| = \left| \sum_i^N \mathbf{j}_i \right|.
\end{equation}
We define the related quantity 
\begin{equation}\label{ang2}
L \equiv \sum_i^N \left|\mathbf{j}_i \right|.
\end{equation}
The dimensionless quantity $| \mathbf{J}|/L$ provides an indication of the extent to which the members of the cluster spin in a common direction. Figure \ref{fig:L_over_L_tilde} plots this ratio as a function of cluster size for two runs at the same temperature. Also plotted are the values of this quantity from constant-density bulk liquid simulations. This is done by evaluating the quantity for atoms within a randomly entered spherical boundary of different sizes. This provides a noise estimate to which the nucleation simulation results may be compared. Across all runs we find that for small clusters sizes ($<10$), the spin is slightly above the noise level, but decays rapidly for larger clusters. Across all simulations, at size $i = 100,$ the ratio is within $| \mathbf{J}|/L < 0.0035 \pm 0.0015$, with the high temperature runs at the high end, and the low at the lower. This is to be expected if the axis of rotation is random relative to the ellipsoidal axis. 

\section{Shapes}
\label{sec:shapes}

One assumption usually made by nucleation models is that the clusters are spherical (with few exceptions, see e.g. Prestipino et al. (2012)  \cite{laio}). 
This is motivated by the sphericity of large liquid droplets, which bear this shape as it minimises the surface area, and therefore the surface energy. While the cluster shapes in our simulations can deviate significantly from any symmetries, for the sake of simplicity, we will investigate the cluster shapes by analysing the extent to which the clusters are ellipsoidal. 
We use principal component analysis to calculate the cluster ellipticities. For each cluster, we calculate the semi-major ellipsoidal axes by following the procedure outlined in Zemp et al 2011\cite{zemp}, which applies this approach to investigate the shapes of dark matter substructure in simulations. We reintroduce this procedure in appendix \ref{PCA_appendix}. Figure \ref{fig:explain_axes} illustrates the results of this analysis for a typical cluster from one of our simulations. 

\begin{figure}
\includegraphics[trim =9cm 8cm 8cm 8.0cm, clip, scale = 0.7]{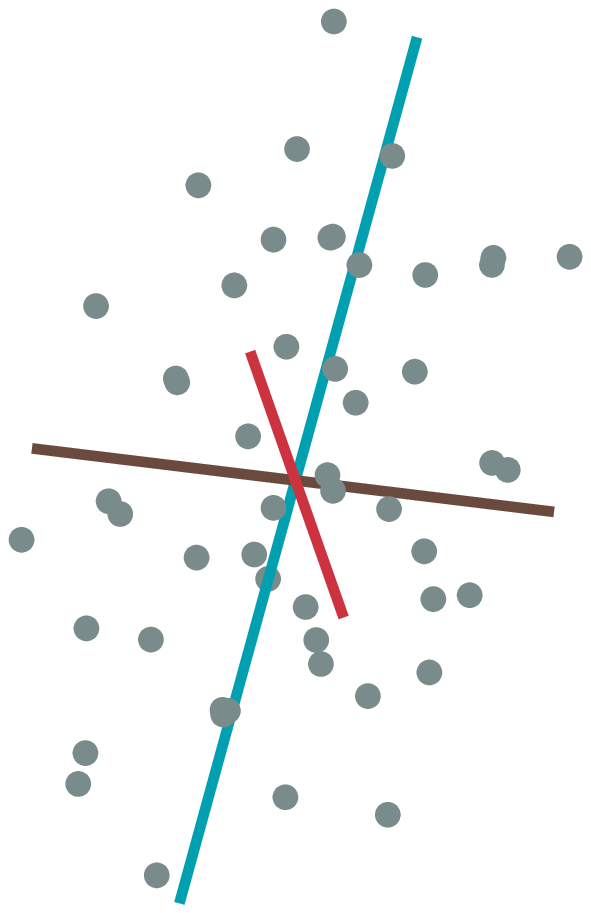}
\caption{The long, medium and short ellipsoidal semi-major axes for a 51-member cluster from T5n40, as acquired from principle component analysis detailed in section \ref{sec:shapes}. This cluster has axis ratios $a/c = 0.45$, and $b/c= 0.63$.}\label{fig:explain_axes}
\end{figure}

 \begin{figure}[]
\includegraphics[scale = 0.6]{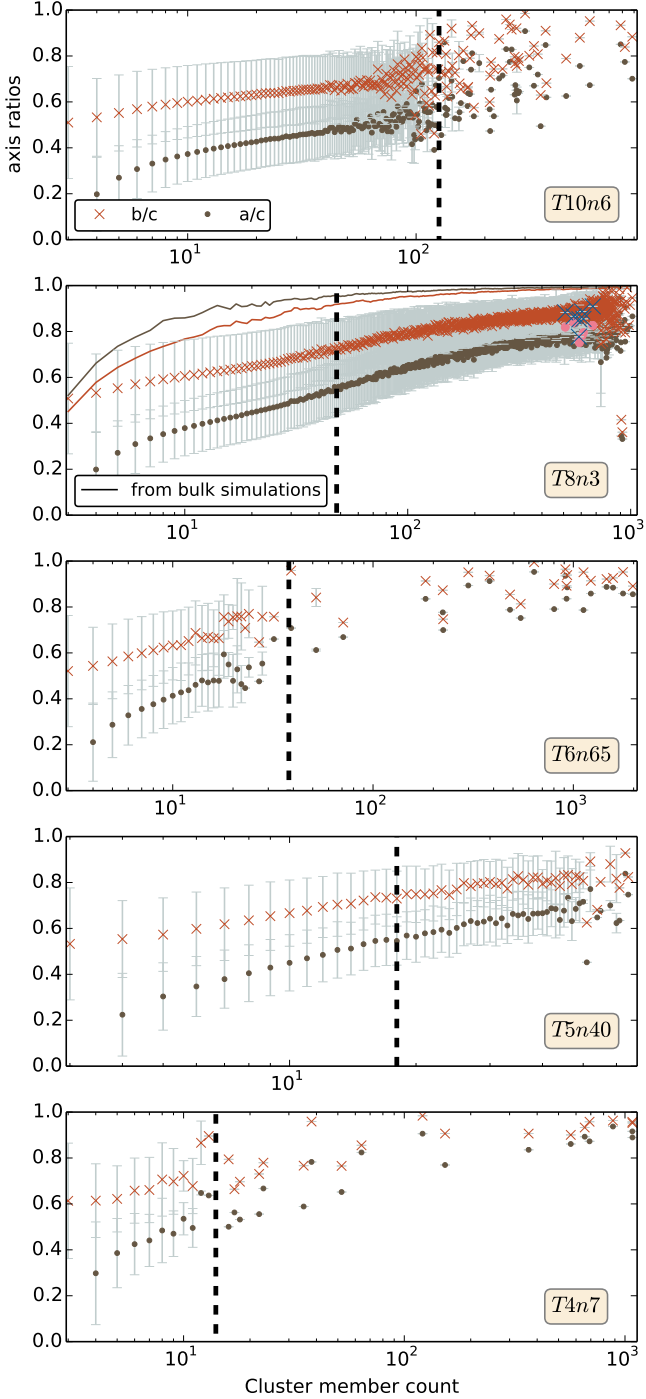}
\caption{Axis ratios as obtained through principal component analysis for 5 runs. Error bars indicate the r.m.s spread. At any given size  high temperature clusters are more elongated than the low temperature clusters. The solid lines in the 2nd panel indicate the axis ratio estimates for 
particle selected within random spheres from a bulk constant density liquid, to illustrate the amount of apparent elongation caused by sampling a sphere with a small number of atoms.}\label{fig:axis_ratios}
\end{figure}

\begin{figure}
\includegraphics[scale = 0.6]{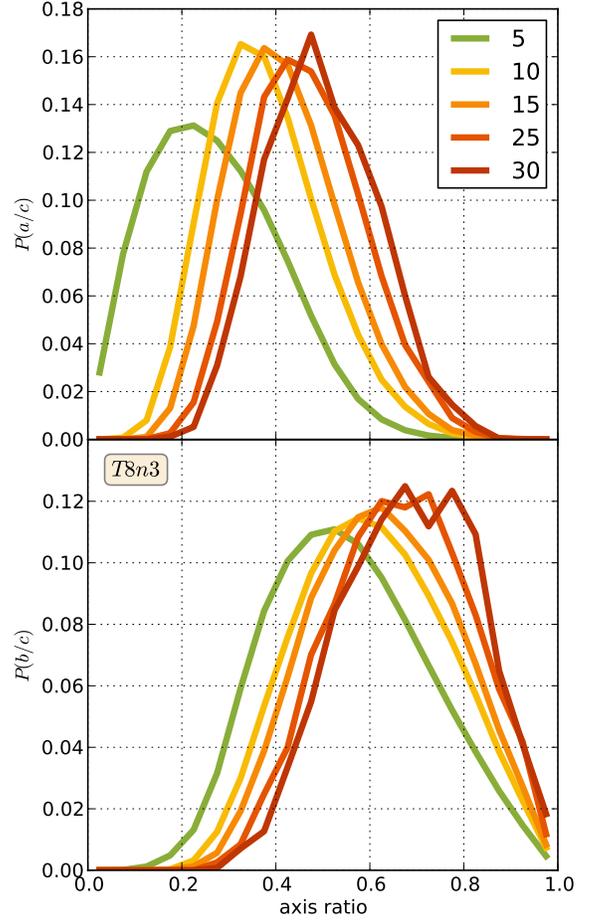}
\caption{The distribution of axis ratios for a single simulation for cluster sizes indicated in the legend inset.}\label{fig:axis_ratio_distribution}
\end{figure}

The cluster axis ratios as a function of cluster size are plotted in figure \ref{fig:axis_ratios}. For a single run, figure \ref{fig:axis_ratio_distribution} shows the probability distribution for the axis ratios, and how they change with cluster size.
 Although the clusters become more spherical as they grow, they are still significantly ellipsoidal at all sizes - in contrast to the standard model assumption that both sub and post-critical clusters are spherical. We observe a trend of increasing ellipticity as temperatures increase. Especially important to nucleation are the shapes around the critical cluster sizes. For each simulation, we find that the critical clusters have axis ratios $b/c = 0.7\pm 0.05$ and $a/c = 0.5 \pm 0.05$. These ellipticities are rather significant, and contrary to standard assumption of spherical shapes in nucleation models.

To explore to origin of these non-spherical cluster shapes we performed supplementary simulations: Eight spherical liquid clusters of size $i = 500$ are placed in a gas at saturation $S = 1$ and $kT = 0.8 \epsilon$. After letting the system equilibrate for 10'000 time-steps, we perform the PCA on the clusters. Their axis ratios are in the second panel (blue crosses and pink circles) of figure \ref{fig:axis_ratios}. We find that they bear the same ellipsoidal distortions as the clusters in the nucleation simulation. We therefore conclude that it unlikely that the clusters' ellipsoidal shapes are in some way dependent on their formation through the nucleation process. Section \ref{sec:rotation} investigates whether angular momentum plays a significant role in cluster dynamics, and finds it relatively insignificant. The relatively low spins suggest that the large ellipticities (figure \ref{fig:axis_ratios}) are not supported by angular momentum.

This leaves us to conclude that the main cause of the cluster ellipticities are thermal fluctuations. High temperatures lead to larger surface fluctuations, and cause larger deviations from sphericity found at high temperatures. Interestingly, the average differences between the axis lengths are nearly independent of cluster size, but increase with temperature: We find that $\overline{a - b} \simeq 0.5\sigma$ and $\overline{a - c} \simeq 1.0\sigma$ for all clusters in the $kT = 0.4 \epsilon$ simulations. Large clusters are rounder (axis ratios closer to one) mainly because their larger size makes the nearly constant absolute differences become smaller in relative terms. The average differences grow with temperature and at $kT = 1.0 \epsilon$ we find $\overline{a - b} \simeq 1.5\sigma$ and $\overline{a - c} \simeq 2.5\sigma$. Short animations of evolving clusters are available, in which their non-sphericity is clearly visible\citep{supp}.

\section{Central Densities and Transition Regions}
\label{sec:densities}

An essential aspect of tiny clusters is the interface layer between the constant-density, liquid interior and the gas outside. The surface energy of the cluster depends on the properties of this layer. Of particular importance to the classical nucleation theory is the surface tension of the droplet, which depends on the interfacial pressure profile. Classical nucleation models assume that clusters are homogeneous, spherical droplets, with a sharp, well defined boundary. At this boundary the fluid properties are assumed to jump from bulk liquid properties on the inside to bulk vapor properties on the outside (see for example Kalikmanov, V.I. (2013)\citep{KalikmanovBook} and Kashchiev (2000)\cite{KashchievBook}). In this section we show that the cluster properties found in our simulation deviate significantly from these assumptions.

Numerical simulations\cite{chapelar} of liquid-vapor interfaces have shown that the density transition is continuous, and for spherical droplets, is well-approximated by
\begin{equation}\label{chapela}
\rho\left(r\right) = \frac{1}{2} \left[\rho_l + \rho_g -\left(\rho_l -\rho_g \right)\tanh \left(2 \frac{r-R}{d} \right) \right],
\end{equation} 
where $\rho_l$ is the density within the cluster, $\rho_g$ the gas density, $R$ the corona position, and $d$ its width. In each cluster's center-of-mass frame, we bin the spherical number density, using a bin size of $0.5 \sigma.$ The number density profiles for clusters of the same size are used to make ensemble averages, to which equation \ref{chapela} can be fit.

\begin{figure}
\includegraphics[scale = 0.65]{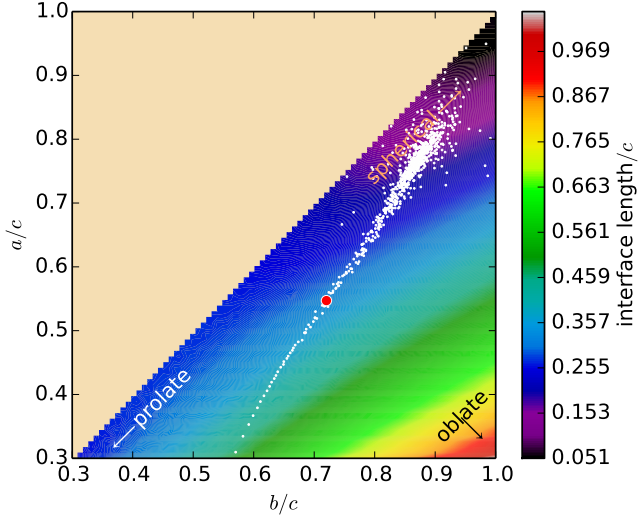}
\caption{The spherical density profile function (\ref{chapela}) relies on the parameter $d$ to characterise the interface length, or, the size of the transition region. Because the clusters are non-spherical, a non-zero interface length can be expected even if the transition is abrupt. This colormap shows the amplitude of this artificial contribution to $d$ as a function of axis ratio. The white dots are the axis ratios for the clusters of T8n3. The red dot marks the critical cluster size.}\label{fig:interface_distortion}
\end{figure}

Because the interface width $d$ is computed from fits to a spherical density profile, yet the clusters do not exhibit spherical symmetry, even a sharp transition region would result in a density profile with $d>0$. Under the more realistic assumption of ellipticity, this artificial contribution to $d$ can be estimated. Figure \ref{fig:interface_distortion} shows the ratio of $d$ and the long axis length $c$, as a function of the axis ratios. We find that this artificial contribution to $d$ is smaller than our measured the interface widths in practically all cases. In other words, this effect, which in principle could lead us to overestimate $d$ significantly, is actually negligible.

\begin{figure}[H]
\includegraphics[scale = 0.6]{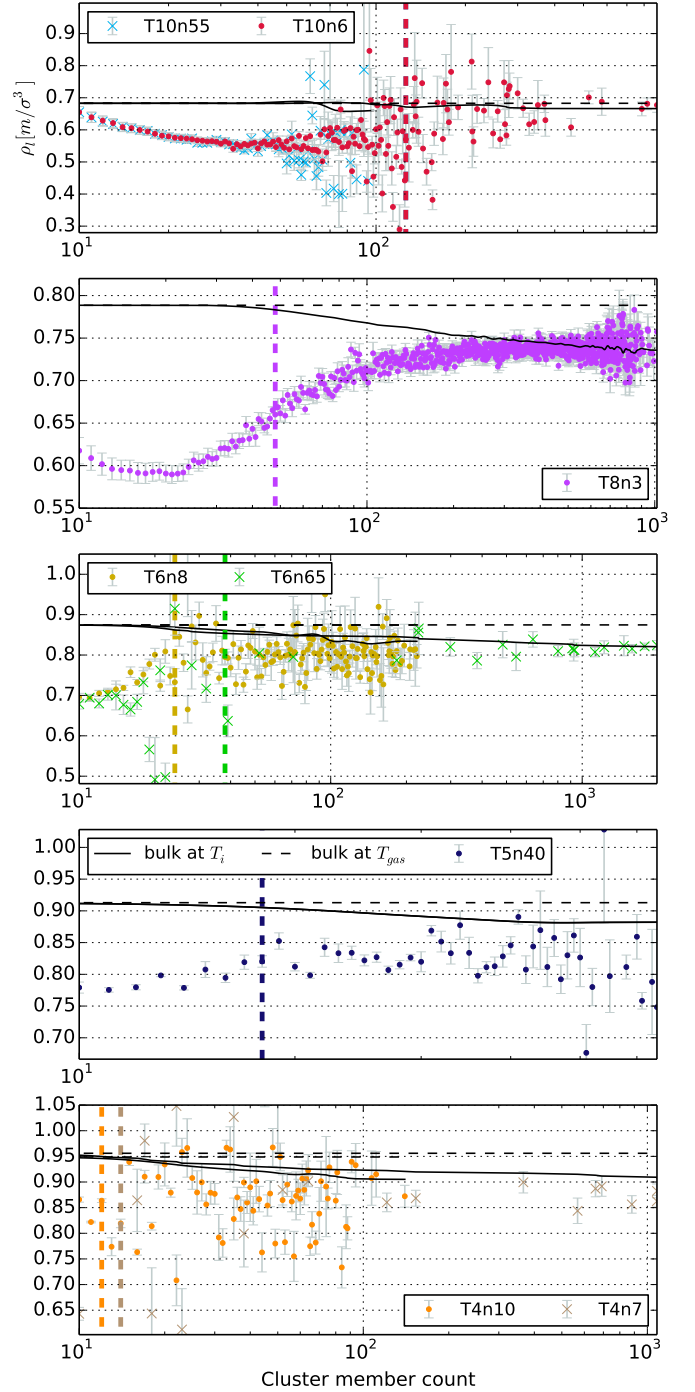}
\caption{Results from the fit parameter for the central density $\rho_l$ in equation (\ref{chapela}). The error bars show the error in the fit. The horizontal dashed black lines are the liquid densities from our bulk simulations, at the gas temperature. The solid black lines represent the bulk density at the running average temperature over size. For clusters much larger than the critical cluster size, the bulk densities at the cluster temperatures agree with the cluster densities. However, there is a significant discrepancy at the critical cluster size of up to 25\%}\label{fig:profile_liquid_density}
\end{figure}

\begin{figure}[!h]
\includegraphics[scale = 0.6]{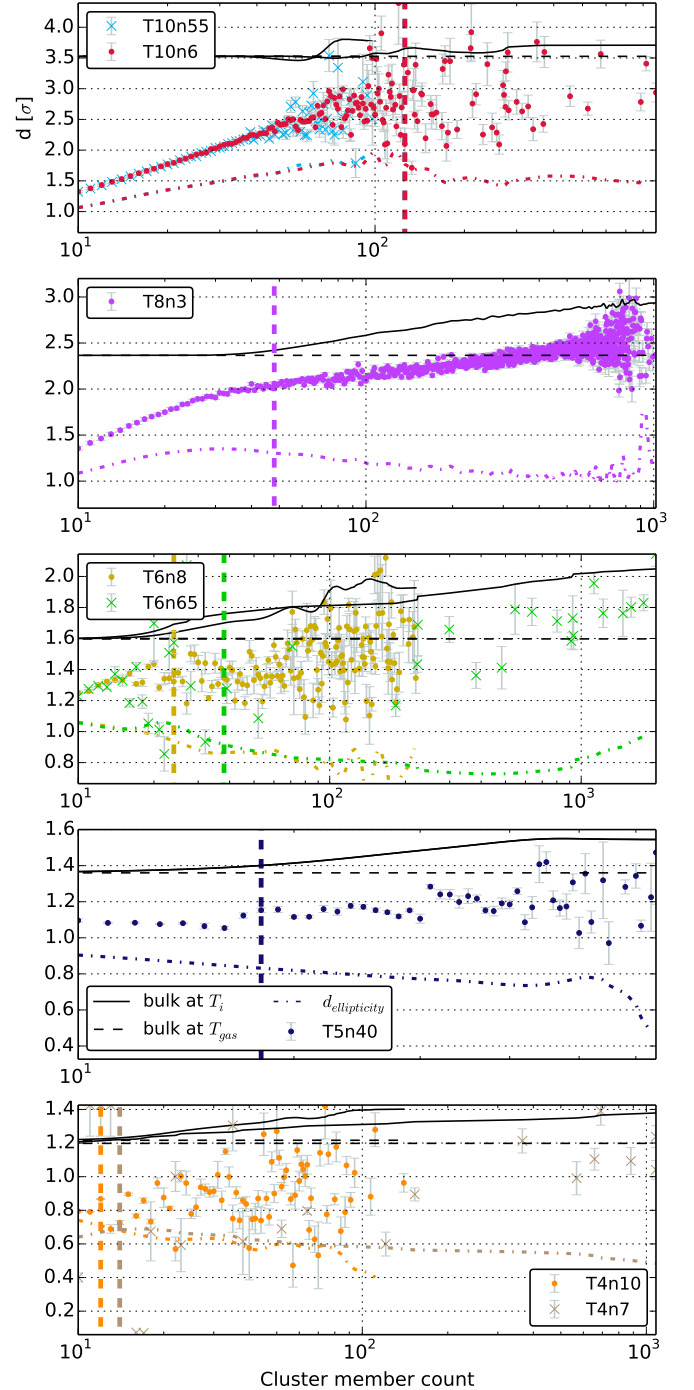}
\caption{ The parameter $d$ (equation \ref{chapela}) is the size of the interphase interfacial region. For clusters of size $i\sim 1000,$ the non-isothermal bulk values overestimate $d$ by $10-30\%$. The dot-dashed curves show $d_{\rm ellipticity}$ - the contribution that the non-spherical shape makes on the determination of $d$ from spherical density profile fits. Refer to figure \ref{fig:interface_distortion} for this quantity's ellipticity dependence. This contribution is always lower than the observed signal.}\label{fig:profile_corona_size}
\end{figure}

The upper panel of Figure \ref{fig:big_cluster_density} shows the binned center-of-mass density profile of the large cluster shown in figure \ref{fig:w00t}, as well as the fit. Figures \ref{fig:profile_liquid_density} and \ref{fig:profile_corona_size} show that the inner density and interface width respectively depend on cluster size: Generally, both the inner density and the interface width increase with cluster size (with the exception of the anomalous inner densities in small, high temperature clusters, see below). Inner densities and interface widths may then be compared to the analogous quantities from equilibrium simulations of planar vapor-liquid interfaces at various temperatures.

We note that:
\begin{itemize}
\item  At small sizes ($i<10$), the clusters hardly have a core, and so equation (\ref{chapela}) gives them only a transition component.
\item At the critical cluster sizes $i=i^*$, the inner densities are significantly lower than in bulk liquid. This implies a surface area larger than expected from classical nucleation theories. 
\item For $i>i^*,$ the clusters are warmer than the gas (see figure \ref{fig:cluster_temperatures}), and due to thermal expansion, have a lower density than the bulk would have at the gas temperature. The bulk densities taken at the cluster temperatures agrees with the central cluster densities only for our very largest clusters ($i>>i^*$). 
\item For all simulations and cluster sizes (i.e. subcritical and post-critical), the cluster transition regions are thinner than the planar. equilibrium interfaces simulated at the same temperature. 
\end{itemize}

Monte-Carlo simulations \cite{tenwolde}, find that critical clusters have inner densities equal to that of bulk liquid, which is at tension with our observations. Napari et al. (2009)\citep{napari} estimated interface widths by comparing the sizes of cluster determined with different cluster definitions. They conclude that critical clusters from direct nucleation simulations have a thicker transition region than spherical clusters in equilibrium with the surrounding vapor. Due to the different simulations and analysis methods a detailed comparison is difficult.

\begin{figure}[]
\includegraphics[scale = 0.6]{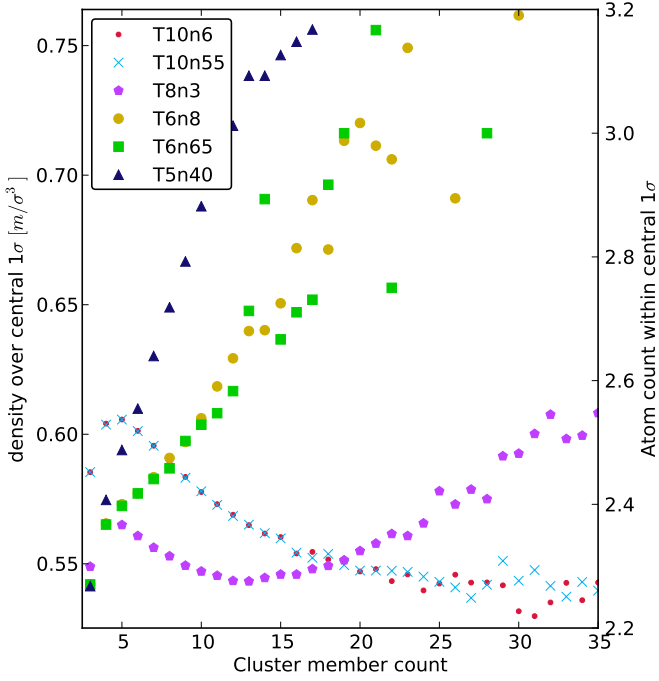}
\caption{Average number of atoms (right scale) and density (left scale) within $1\sigma$ of the cluster centre of mass. Generally, the central densities increase with size and our largest clusters reach the bulk liquid value (see figure \ref{fig:profile_liquid_density}).
However, at high temperature ($kT = 0.8, 1.0\epsilon$), we observe a surprising anomaly: the central densities drop and reach a well defined minumum before they rise again.}\label{fig:central_density}
\end{figure}

As mentioned above, figure \ref{fig:profile_liquid_density} shows that the inner densities generally increase with cluster size, except for small, high temperature clusters. Fitting the density profile of small clusters is difficult, because they do not have a well defined, constant inner density. This could affect the resulting
$\rho_l$ values. To confirm the surprising anomaly in the inner densities of small, high temperature clusters we also measure the central cluster densities within a $1\sigma$ sphere of the cluster centre of mass directly:  
Figure \ref{fig:central_density} plots the ensemble-averaged number of particles within $1\sigma$. This alternative measure confirms the findings from $\rho_l$ directly, without any fitting procedure: Generally, as the clusters grow, they become more tightly bound, leading to an increased central density. The small clusters in our high temperature runs at $kT = 0.8, 1.0\epsilon$ show a different trend: the central density first decreases, and then increases again. At $kT=0.8\epsilon,$ the minimum central density occurs at $i = 13$, and at $kT = 1.0\epsilon,$ it occurs in the range $30 \leq i \leq 40.$
This anomaly is evident both in the $\rho_l$ values from the fits and in the densities within the central $1 \sigma$. We are currently unable to explain this behaviour.

\section{Cluster sizes}
\label{sec:sizes}

\begin{figure}[]
\includegraphics[scale = 0.6]{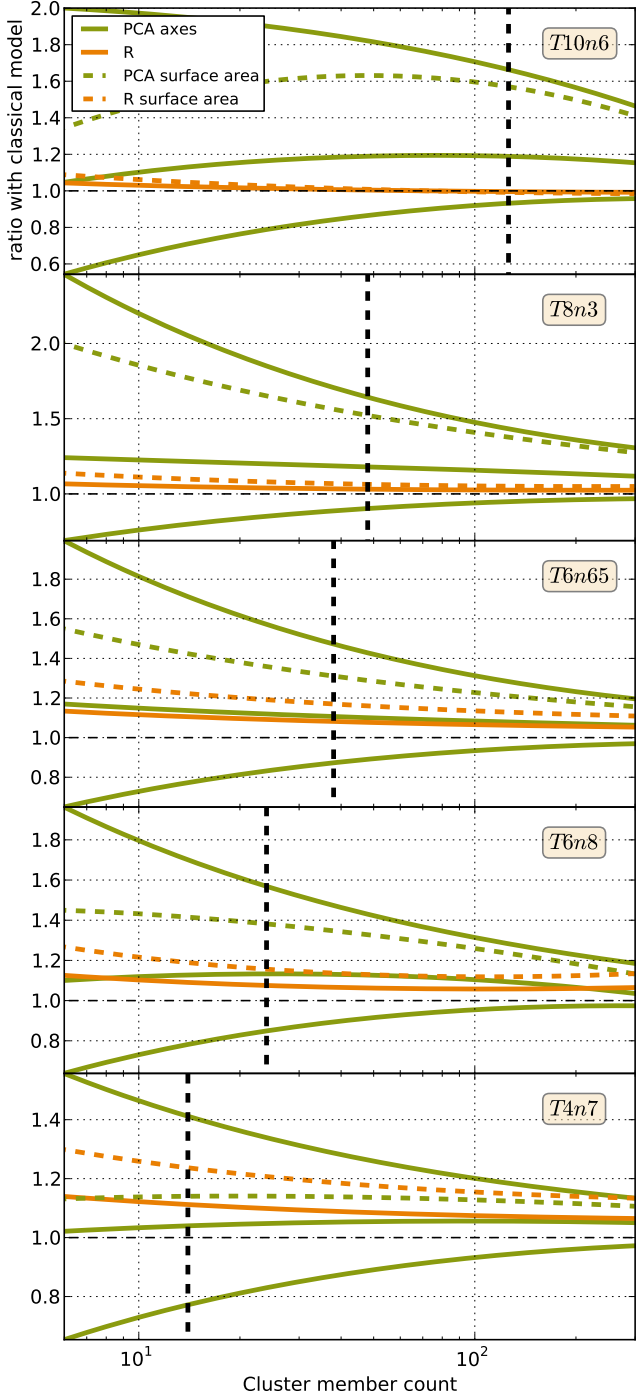}
\caption{The solid green lines show the axes sizes (from principle component analysis) relative to the bulk expectation. Fitting functions were used to compute all the ratios shown here. The orange depict $R$ from spherical density profile fits (equation \ref{chapela}), relative to the bulk value. The dashed lines are the surface areas corresponding to these two size estimates. PCA overestimates the sizes significantly at small sizes. Density profile fits give a more conservative value for the sizes and surface areas, yet still significantly larger than nucleation models' predictions.}\label{fig:sizes}
\end{figure}

We have two means for measuring cluster sizes. The one is with the principle component analysis procedure, detailed in section  \ref{sec:shapes} and appendix \ref{PCA_appendix}, and the other with density profile fits, explained in section \ref{sec:densities}. The principle component analysis route, under the assumption of a constant density ellipsoid, yields the three ellipsoidal axes $a, b $ and $c$ for each cluster. The second method assigns to each density profile the centre of the transition region, $R$. For nucleation, sizes are important because they provide an estimate for the cluster surface area, which helps set the total surface energy - a key component for nucleation theories. The simplest nucleation models assume that a cluster with $i$ members is spherical, and has a density equal to that of the bulk at the same temperature, from which a size, and therefore surface area, can be calculated. For three ellipsoidal axes $a, b$ and $c$, the surface area may be analytically obtained using the approximate relation \cite{ellipse}
\begin{equation}
S_{\rm ellipsoid} \approx 4\pi\left[\frac{a^p b^p + a^p c^p +b^p c^p}{3} \right]^{\frac{1}{p}} \;\;\;\; \textrm{with} \;\;\;\; p =1.6075, 
\end{equation}
which has a worst-case relative error $< 1.061\%$. Figure \ref{fig:sizes} compares, using our two size-measuring methods, the sizes and surface areas of the clusters to the standard nucleation model assumptions.

Both methods give larger sizes and surface areas than the classical model assumptions, due to the densities being lower than the bulk. Both methods suffer from large uncertainties and it is unclear how their resulting surface areas relate to the area of the true (unknown) surface of tension. The surface of tension lies somewhere in the transition regions. We have tried to calculate the radius of the surface of tension and the surface energy by assuming spherical interfaces and using the Irving-Kirkwood\citep{KI} pressure tensor approach applied to spherical droplets\citep{Thompson1984, song}. However, the transient nature of the non-equilibrium clusters in our nucleation simulations does not allow for the accumulation of strong-enough statistics for us to get a useful surface energy signal. We are therefore unable to constrain the cluster surface tension and the location of their true surface of tension. 

For critical clusters we observe $d/R$ ratios from about 0.6 to 0.85 (see Figure \ref{fig:profile_corona_size}). This introduces very large possible shifts in the surface areas: Placing the surface of tension at $R+d/2$ instead of $R$ would increase the critical cluster sizes by 30 to 43 percent, and their surface areas by factors of 2.2 to 2.9. Setting the surface of tension to $R-d/2$ instead of $R$ would decrease surface areas by factors of 2.9 to 5.3. These examples illustrate how large the uncertainties in the area of the surface of tension are, which introduces huge uncertainties of many orders of magnitudes into any nucleation rate predictions based on these surface areas.

Compared with the spherical density profile based size definition, which uses the midpoint of the transition region  $R$ as cluster radius, the principle component analysis method usually gives larger sizes. This is because of the assumption that clusters are constant density ellipsoids, when converting the eigenvalues to axis lengths using equation \ref{abc}. The PCA analysis weights outer members heavily in the computation, yet these outer members are not part of a constant density neighbourhood, because they belong to the tail of the transition region. This effect decreases as the size of the transition region becomes small relative to the cluster size, i.e. when $d/R \rightarrow 0$. At low temperatures, the PCA route yields smaller clusters than the density profile method. This, because low-temperature clusters are more spherical than higher temperature ones.


\section{Revisiting nucleation models}\label{sec:revisiting}

\begin{figure}[]
\includegraphics[scale = 0.6]{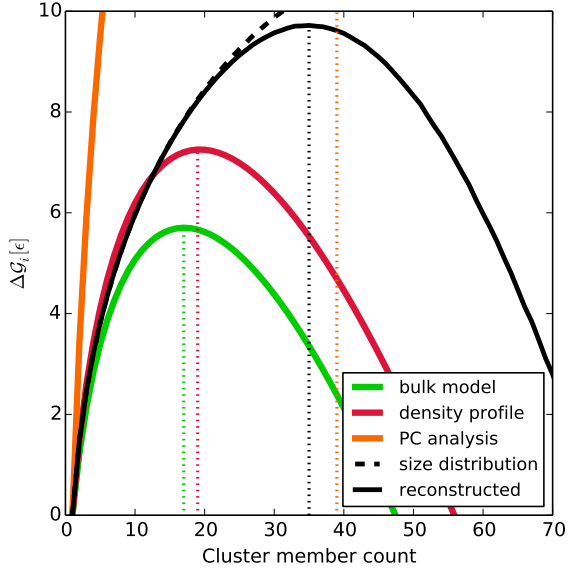}
\caption{The solid lines are free energy curves for the run T8n3, each using different estimates for the surface area (see figure \ref{fig:sizes}). All curves use the semi-phenomenological model for the surface tension. The lower-than-expected bulk densities lead to more significant surface terms (equation (\ref{funtimes})). The black dashed line is the equilibrium free energy component, calculated from the cluster size distribution, and the solid is the reconstructed free energy [Tanaka et al. (2014)]. The bulk, density profile, and PCA curves correspond to nucleation rates $3.5\cdot10^{-7}$, $5.2\cdot10^{-8}$, and $9.7\cdot10^{-17}$ $ \sigma^{-3}\tau^{-1}$ respectively. The directly-measured MD value lies at $ 2.6 \cdot 10^{-10} \sigma^{-3}\tau^{-1}$.}\label{fig:delG_it8n3}
\end{figure}

\begin{figure}[]
\includegraphics[scale = 0.6]{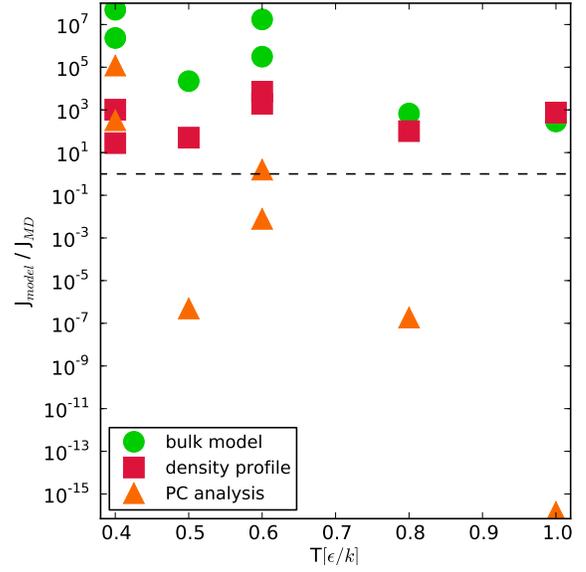}
\caption{Using the SP model for the cluster surface tension, and our three different surface area estimates, we can construct free energy curves like those in figure \ref{fig:delG_it8n3} for each simulation, and compare the nucleation rates they correspond to, with those we measure directly from the simulations. }\label{fig:models_compare}
\end{figure}

Nucleation models for the free energy of formation aspire to find the balance between the energy gain and cost due to creation of volume and surface.
The volume term is well-understood as its contribution to the formation energy dominates in the large-cluster limit, whose properties are therefore straightforward to verify. Nucleation models' shortcomings are thought to be due to an insufficient understanding of the surface energy contribution to the free energy, which dominates for small clusters, and which is therefore difficult to verify. Most nucleation models in the literature offer various forms for the surface energy component. 
 For example,  a common choice for the surface energy expresses the surface tension of a spherical cluster as a correction to the planar surface tension. This prescription for the free energy takes the form \citep{oxtoby, laaky-waaky}
\begin{equation} \label{funtimes}
\Delta G_i = \underbrace{- i\, kT\ln S}_{\textrm{volume energy}} + \underbrace{\underbrace{\underbrace{\gamma_\infty}_{\textrm{planar value}} \underbrace{ \left(1- \frac{2\delta}{r_i}  + \frac{\epsilon}{r^2_i}\right)}_{\textrm{curvature correction}}}_{\textrm{surface tension}} \underbrace{4\pi r_i^2}_{\textrm{surface area}}}_{\textrm{surface energy}},
\end{equation}
where $r_i = i^{1/3}r_0$ is the cluster radius.
The Dillman-Meier approach\citep{SPmodel} lets the $\delta$ term play the role of a Tolman-like \cite{tolman} correction. The SP model as used in Tanaka et al. \citep{tanaka2011} and Diemand et al. \citep{paper1} makes the choice
\begin{equation}\label{SP}
\delta = -\frac{kT}{8\pi\gamma_{\infty}r_0}\xi, \;\;\;\;\;\; \epsilon = 0,
\end{equation}
where $\xi$ is set using the second virial coefficient, and
\[
r_0 = \left(\frac{3}{4\rho\pi}\right)^{1/3},
\] 
with the density $\rho$ taken to be equal to that of the bulk density. The classical nucleation model on the other hand, lets the surface tension take on the planar value, regardless of cluster size, setting $\delta = \xi =\epsilon = 0$. Both the CNT and SP models however, along with many others, make the same choice for the surface area, setting it to $4\pi r_i^2.$ In this section we explore the effect of replacing this surface area estimate with the directly-measured values 
\begin{enumerate}
\item from principal component analysis (\ref{sec:shapes}), and
\item from density profile fits (\ref{sec:densities}).
\end{enumerate}
For the surface tension, we use the SP model parameters (\ref{SP}). We impose the further stipulation that the free energy of formation for a cluster of size one is zero:
\begin{equation}
\Delta G_i \rightarrow \Delta \mathcal{G}_i = \Delta G_i - \Delta G_1.
\end{equation}
Figure \ref{fig:delG_it8n3} plots modelled free energy curves for a single run, and compares them to a kinematically reconstructed  (Tanaka, K. et al (2014), in preparation) free energy. Our techniques for estimating the sizes (see section \ref{sec:sizes} and figure \ref{fig:sizes}) show that because the cluster densities are lower than the bulk values, the surface areas are larger than the traditional nucleation model surface area assumptions. This increases the cost in forming a surface, lowering nucleation rates. Figure \ref{fig:models_compare} compares the resultant nucleation rates to those measured directly from the simulation using the Yasuoka-Matsumoto\citep{yasuoka} (or threshold) method. In almost all cases, the directly-measured surface areas lower the nucleation rate. The density profile surface area estimates improve the model estimation by a factor $10$ - $10^5.$ However, the PCA surface area estimates significantly underestimate the nucleation rates, especially at high temperature, where surface fluctuations dominate the size measuring method. One should not keep ambitions for retrieving perfect-agreement nucleation rates with the procedure used in this section, as our `empirical' surface energy model is still at the behest of a theoretical surface tension model, which likely holds shortcomings unfortunately typical to droplet surface tension models. Given that reasonable size measurements improve the model predictions, we are lead to conclude that it is not just the surface tension modelling which needs improvement - but that models must address surface area estimates directly, taking into account the not-yet-bulk central densities of clusters.

\section{Conclusions}

This work offers detailed description of cluster formation in unprecedented large-scale Lennard-Jones molecular dynamics simulations of homogeneous vapor-to-liquid nucleation. Our main findings are

\begin{itemize}
\item Significant latent heat is retained by large, stable clusters: as much as $\Delta kT = 0.1 \epsilon$ for clusters with $i = 100$. Small, sub-critical clusters on the other hand have the exact same average temperature as the surrounding vapor.
\item Cluster shapes deviate significantly from spherical: ellipsoidal axis ratios for critical cluster sizes lie typically within $b/c = 0.7\pm 0.05$ and $a/c = 0.5 \pm 0.05$. 
\item Cluster spin is small and plays a negligible role in the cluster dynamics.
\item For critical, sub-critical, and post-critical clusters, the central potential energies per particle are significantly less negative than in the bulk liquid. They reach the bulk values only at large ($i>100$) sizes.
\item Central cluster densities generally increase with cluster size. However, for small, high temeperature clusters we uncover a surprising exception to this rule: their central densities decrease with size, reach a minimum (at $i = 13$ for $kT=0.8\epsilon$ and at $i \simeq 35$ for $kT = 1.0\epsilon$) and then join the general trend of increasing central densities with larger sizes.
\item For critical and sub-critical clusters, $i \le i^*$, the central densities are significantly smaller than in the bulk liquid. At the critical cluster size, the cluster central densities are between $5-30\%$ lower than the bulk expectations. This implies a surface area larger than expected from classical nucleation theories. 
\item For $i>i^*,$ the clusters are warmer than the gas (see figure \ref{fig:cluster_temperatures}), and due to thermal expansion, have a lower density than the bulk would have at the gas temperature.
The bulk densities and potential energies taken at the cluster temperatures agree with the central cluster properties only for our very largest clusters ($i>>i^*$). 
\item For all simulations and cluster sizes (i.e. subcritical and post-critical), the cluster transition regions are thinner than the planar equilibrium interfaces simulated at the same temperature. 
\item Cluster size measurements suggest larger sizes than assumed in classical nucleation models, implying lower-than-expected nucleation rates. However, the exact area of the true surface of tension remains unknown, as does the surface tension itself - a major source of uncertainty in nucleation rate predictions.
\item Across all the cluster properties examined in this paper, there exists significant spread for clusters at each size. The standard approach to nucleation theory assumes   all clusters of the same size have the same properties. This therefore allots to all clusters of a certain size, the same surface energy, as opposed to distributing them into disparate population types. The observed scatter in the cluster properties at each size suggests that the development of nucleation theories which address this may lead to a more realistic description of the process. 
\end{itemize}

\begin{acknowledgments}
We thank the referees for useful comments. We acknowledge a PRACE award (36 million CPU hours) on Hermit at HLRS. 
Additional computations were preformed on SuperMUC at LRZ, on Rosa at CSCS and on the zBox4 at UZH.
J.D. and R.A. are supported by the Swiss National Science Foundation.
\end{acknowledgments}

\appendix

\section{Determining cluster shapes with principal component analysis}\label{PCA_appendix}
We define the tensor 
\begin{equation}
\mathbf{M} \equiv \int_V \rho\left(\mathbf{r}\right) \mathbf{r} \mathbf{r}^{T} \textrm{d} V,
\end{equation}
which is the second moment of the mass distribution - the part of the moment of inertia tensor responsible for the describing the matter distribution. The shape tensor is defined by
\begin{equation}
\mathbf{S} \equiv \frac{\mathbf{M}}{M_{\textrm{tot}}} = \frac{\int_V \rho\left(\mathbf{r}\right) \mathbf{r} \mathbf{r}^{T} \textrm{d} V}{\int_V \rho\left(\mathbf{r}\right)\textrm{d} V}.
\end{equation}
For discrete, equal mass particles, in the centre-of-mass frame, the elements of the shape tensor read 
\begin{eqnarray}
S_{\textrm{C.o.M.}, kj} &=& \frac{1}{N} \sum_i^N \left( \mathbf{r}_i \right)_k \left( \mathbf{r}_i \right)_j \\
&=&\frac{1}{N} \sum_i^N 
\left( \begin{array}{ccc}
x^2_i & x_i\cdot y_i & x_i\cdot z_i \\
x_i \cdot y_i & y^2_i & y_i\cdot z_i \\
x_i\cdot z_i & y_i\cdot z_i & z^2_i \end{array} \right),
\end{eqnarray}
in cartesian coordinates relative to the cluster centre of mass, where the sum is over all the members of the cluster. 
If there exist vectors $\mathbf{V}_l$, for $l = a,b,c$ which satisfy 
\begin{equation}
  \mathbf{S}_{\textrm{C.o.M}} \mathbf{V}_l  =  \lambda_l \mathbf{V}_l ,
\end{equation}
then the triplet  $\mathbf{V}_l$ and $\lambda_l$ form the eigenvector-eigenvalue pairs for $ \mathbf{S}_{\textrm{C.o.M}}$. For an ellipsoid of uniform density, the axes $a, b$ and $c$ are related to the eigenvalues of the shape tensor via 
\begin{equation}\label{abc} 
a,b,c = \sqrt{3 \lambda_{a,b,c} }.
\end{equation}
We choose the convention $a<b<c$. Clusters not large enough to have a significant number of core atoms - therefore composed only of a fluffy surface, cannot be well-approximated by a constant-density ellipsoid. For these clusters, the method can overestimate the axis lengths, implying that this method does not provide a sound estimate of clusters' sizes when they are small. However, the axis ratios provide a useful indicator of the cluster shapes regardless of the cluster size. 

\section{Supplementary equilibrium simulations of planar vapor-liquid interfaces}\label{bulk}

To determine the thermodynamic properties of the Lennard-Jones fluid simulated here a range of liquid-vapor phase equilibrium simulation were run, similar to those in \citep{slabz, baidakov, Trokhymchuk, Mecke1997}. We used the same setup and analysis as described in \citep{baidakov}, except that we use a different cutoff scale ($r_{\rm cut} = 5 \sigma$) and a different time-step ($\Delta t = 0.01 \tau$). For simplicity, we calculate the surface tensions using the Kirkwood-Buff pressure tensor only and to obtain better statistics the interface surface area was increased by doubling the size of the simulated rectangular parallelepiped in the x and y direction, i.e. we set $L_x \times L_y \times L_z = 27.12 \sigma \times 27.12 \sigma \times 58 \sigma$.

\begin{figure}
\includegraphics[scale = 0.6]{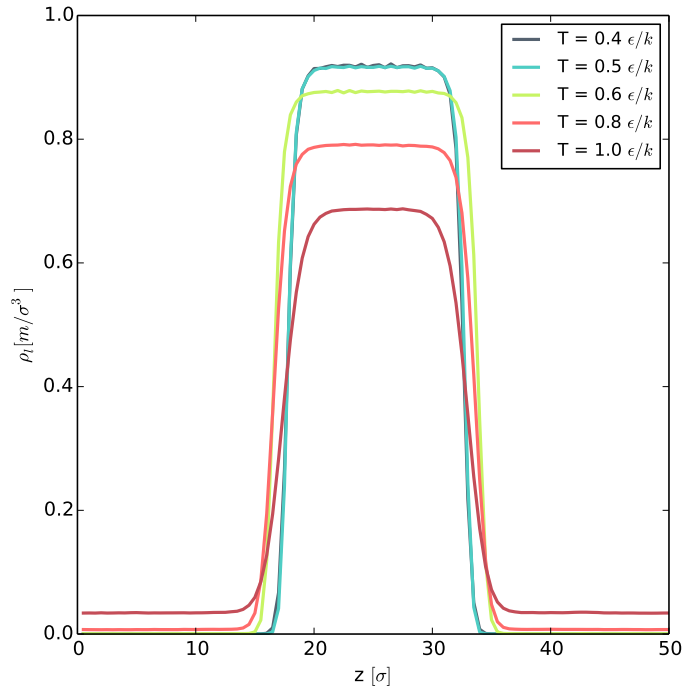}
\caption{Density profiles from rectangular box bulk simulations, detailed in appendix \ref{bulk}.}\label{fig:bulk_profiles}
\end{figure}

\begin{figure*}
\includegraphics[scale = 0.6]{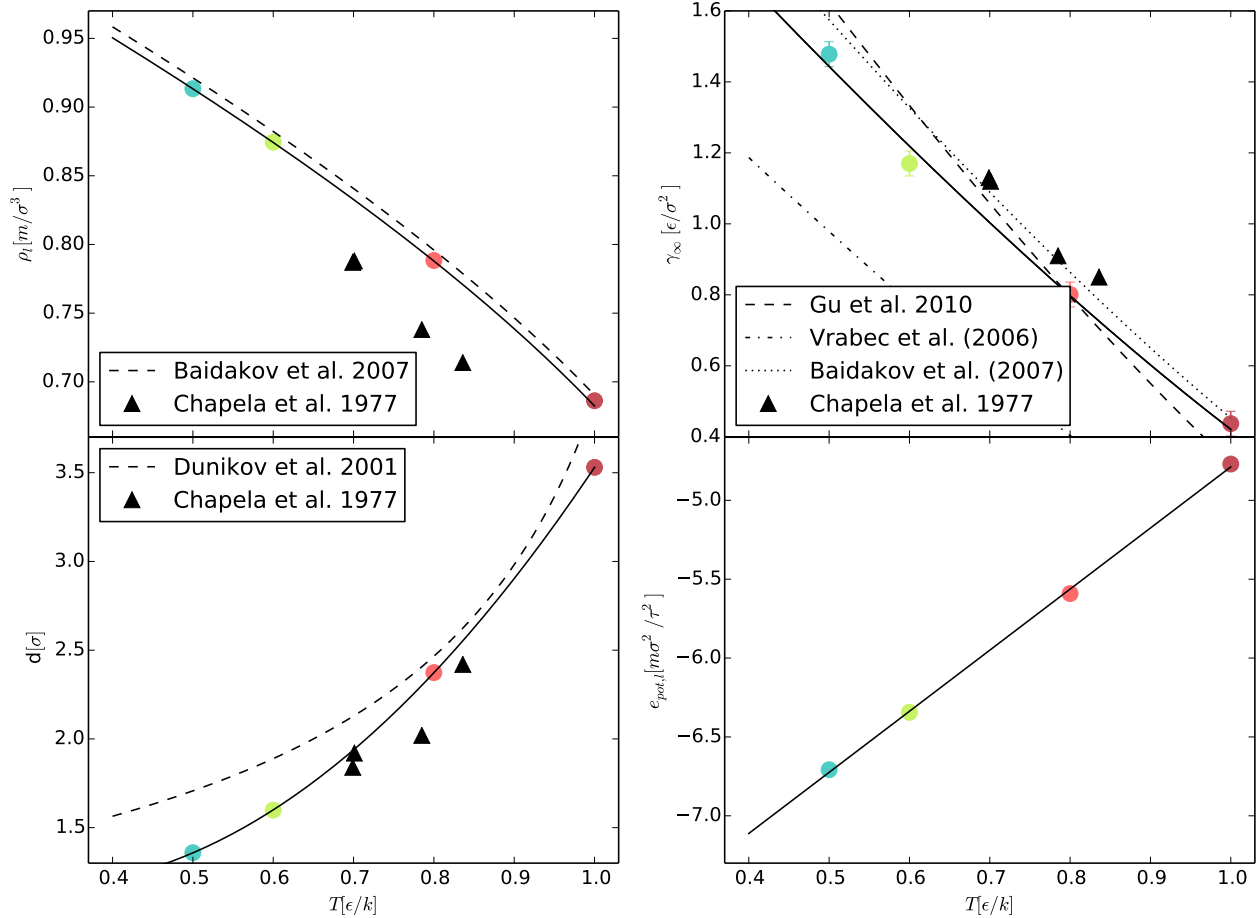}
\caption{Thermodynamic parameters from equilibrium simulations of planar vapor-liquid interfaces. Literature values \citep{baidakov, gu, dunikov, chapelar, moargamma} are shown for comparison. Differences in the measured parameters can be attributed to a number of factors, the choice for the cutoff-scale and the simulation size foremost among them: For example, Vrebec et al (2006)\citep{moargamma} truncate the Lennard-Jones potential at $r_{\textrm{cut}} = 2.5 \sigma$, Baidakov et al. (2007)\citep{baidakov} use $r_{\textrm{cut}} = 6.7 \sigma$ while Chapela et al. (1977)\citep{chapelar} have $N\sim 10^3$.}\label{fig:bulk_parameters}
\end{figure*}

Using the same cutoff scale ($r_{\rm cut} = 6.78 \sigma$) as in \citep{baidakov}, we are able to exactly reproduce their results given at $T= 0.5, 0.6. 0.8$ and $1.0 \epsilon / k$.
The results with our chosen cutoff scale ($r_{\rm cut} = 5 \sigma$) are shown in Figure \ref{fig:bulk_profiles}.
The surface tensions $\gamma$ are about 5 percent lower, and the bulk liquid densities lower by $\sim0.01 m/\sigma^3$ than those found in \citep{baidakov}. Figure \ref{fig:bulk_profiles} also shows fitting functions to our simulation results for
the bulk liquid density
\begin{equation}\label{eqn:rhol}
\rho_m = 0.0238\cdot \left(13.29 + 24.492 f^{0.35} + 8.155 f\right) - 0.008 m/\sigma^3,
\end{equation}
with 
\begin{equation}
f = 1-\frac{kT}{1.257}
\end{equation}
the planar interface thickness
\begin{equation}\label{eqn:d_infty}
d_\infty = -2.87\cdot kT + 4.82\cdot kT^2 + 1.59,
\end{equation}
the planar surface tension
\begin{equation}\label{eqn:gamma_infty}
\gamma_\infty = 2.67 \times \left(1 - T / T_c \right)^{1.28} \epsilon/\sigma^2\; , T_c = 1.31 \epsilon/k, 
\end{equation}
and the potential energy per particle in the bulk liquid
\begin{equation}\label{eqn:epotbulk}
e_{\rm pot,l} = 3.872\cdot kT - 8.660.
\end{equation}
We use the values of these fitting functions throughout this paper, the values at $T= 0.4, 0.5, 0.6, 0.8$ and $1.0 \epsilon / k$ are listed in table \ref{tab:bulksims}. Results of the thermodynamic quantities that we calculate from slab simulations, and comparison to similar simulations by other authors are plotted in figure \ref{fig:bulk_parameters}.

Note that at $T= 0.4 \epsilon / k$ we have to rely on extrapolations . We could not get meaningful constraints from equilibrium simulations at this low very temperature, because our liquid slabs begin to freeze before a true equilibrium with the vapor is established. The same limitations were also reported in Baidakov et al. (2007)\citep{baidakov}.

The saturation pressures $P_{\rm sat}$ in our equilibrium simulations agree well with the fitting function proposed in Trokhymchuk et al.\cite{Trokhymchuk}, and we
use this fitting function and the actual pressure $P$ measured in the nucleation simulations to determine the supersaturation $S = P / P_{\rm sat}$, see Diemand et al 2013\citep{paper1} for details.

\begin{table}[]
\caption{Planar surface tensions $\gamma_\infty$, bulk liquid densities $\rho\sub{l}$ and potential energies per atom in the bulk liquid $e_{\rm pot,l}$ from fits to our vapor-liquid equilibrium simulations, detailed in appendix \ref{bulk}.
$\xi$ are from Diemand et al 2013\citep{paper1}.}
\begin{ruledtabular}
\begin{tabular}{l c c c c}\label{tab:bulksims}
 $T$  & $\rho_l$  & $\gamma_\infty$ & $e_{\rm pot,l}$ & $\xi$\\
 $[\epsilon/k]$ & $[m/\sigma^3]$ & $[\epsilon/\sigma^2]$ & $[m\sigma^2/\tau^2]$& \\\hline
 1.0 & 0.682 & 0.437 & -4.77 & 1.94\\\hline
 0.8 & 0.787 & 0.801 & -5.591 &-1.52\\\hline
 0.6 & 0.874 & 1.17 & -6.344 &-6.21\\\hline
 0.5 &0.913 & 1.48 & -6.708 &-9.46\\\hline
 0.4 & 0.950 & 1.57 & -7.10 &-13.9\\
\end{tabular}
\end{ruledtabular}
\end{table}



%



\end{document}